\documentclass[%
 reprint,
showpacs,preprintnumbers,
 amsmath,amssymb,
 aps,
]{revtex4-1}

\renewcommand{\P}{\mathcal{P}}
\newcommand{\bt}{\pmb{\theta}}
\newcommand{\bl}{\pmb{\lambda}}
\newcommand{\bL}{\pmb{\Lambda}}
\newcommand{\bh}{\pmb{\text{h}}}

\DeclareMathOperator*{\argmax}{arg\,max}

\usepackage{hyperref}
\usepackage{color}
\usepackage{graphicx}
\usepackage{algorithmic}
\usepackage{dcolumn}
\usepackage{bm}

\begin{document}

\preprint{PRL}

\title{Statistical Inference for Valued-Edge Networks:\\ Generalized Exponential Random Graph Models.}

\author{B. A. Desmarais$^1$ and S. J. Cranmer$^2$}
\affiliation{$^1$Department of Political Science, University of Massachusetts at Amherst, Amherst, MA 01003\\
$^2$Department of Political Science, University of North Carolina at Chapel Hill, Chapel Hill, NC 27599
}%

\date{\today}

\begin{abstract}
\noindent Across the sciences, the statistical analysis of networks is central to the production of knowledge on relational phenomena. Because of their ability to model the structural generation of networks, exponential random graph models are a ubiquitous means of analysis. However, they are limited by an inability to model networks with valued edges. We solve this problem by introducing a class of generalized exponential random graph models capable of modeling networks whose edges are valued, thus greatly expanding the scope of networks applied researchers can subject to statistical analysis.
\end{abstract}


\maketitle


The need to analyze networks statistically transcends disciplines that have occasion to study the relationships between units.  
Applications in physics \cite{Karrer:2010, Karrer:2009, Newman:2009a, Garlaschelli:2004,Bianconi:2001}, computer science \cite{Myers:2010}, the social sciences\cite{Butts:2008, Cranmer:2011}, and other fields examine networks that vary in size and density, over time, and have edges with values that vary from binary ties, to counts, to bounded continuous and unbounded continuous edges. An important method for statistical inference on networks is the exponential random graph model (ERGM)\cite{Holland:1981,Berg:2002, Park:2004}, which estimates the probability of an observed network conditional on a vector of network statistics that capture the generative structures in the network. 
Yet the ERGM has a major limitation: it is only defined for networks with binary ties\cite{Robins:1999,Wyatt:2010}, thus excluding a wide range of networks with valued edges (e.g., gene co-expression networks, passage time on networks of various media, monetary transactions, casualties in conflict networks).

We develop a class of generalized ERGMs (GERGMs) for inference on networks with continuous edge values, thus lifting the restriction of this methodology to a, possibly small, subset of networks. 
The form of our generalized model is similar to the ERGM in that it can be flexibly  specified to cover a broad range of generative features. The GERGM can be estimated efficiently with a Gibbs sampler.  

The strengths and limitations of the ERGM are apparent from its specification. Let $Y$ be the $n$-vertex network (adjacency matrix) of interest with $m$ edges ($m = n(n-1)$ if $Y$ is directed and $n(n-1)/2$ if it is undirected). $Y_{ij}$ is the edge from $i$ to $j$.  An ERGM of that network is specified as:
\begin{equation}
\P(Y, \bt) =\frac{\exp \{ \bt'\hspace{2pt} \bh(Y) \}}{\sum_{\text{all }Y^* \in \mathcal{Y}} \exp \{\bt'\hspace{2pt}\bh(Y^*) \}},
\label{ergm} 
\end{equation}  
where $\bt$ is a parameter vector, $\bh(Y)$ is a vector of statistics on the network, and the object of inference is the probability of the observed network among all possible permutations of the network given the network statistics. The $\bh(Y)$ term is what gives the ERGM much of its power: this vector can contain statistics to capture the endogenous structure of connectivity in the network (statistics can be included to capture reciprocity, transitivity, cyclicality, and a wide variety of other endogenous structures) as well as the effects of exogenous covariates.

The challenges for modeling networks with valued edges are apparent from the specification in equation \ref{ergm}. The flexibility of the distribution comes from the lack of constraints in specifying $\bh$; the only constraint is that $\bh$ is finite when evaluated on any binary network. This assures that the denominator is a \emph{convergent} sum, and therefore represents a proper normalizing constant for the distribution of networks. However, this convergence is not assured whenever $\bh$ is finite if the support of $Y$ is infinite. The model we derive retains the flexibility of $\bh$ within a framework that assures a proper probability distribution for $Y$ when $Y$ has continuous edges.

Our generalized ERGM operates by constructing joint \emph{continuous} distributions on networks that permit the representation of dependence features among the elements of $Y$ through a set of statistics on the network, $\bh(Y)$. 
As in the ERGM, the vector $\bh$ can be specified to represent many forms of dependence, including transitivity (i.e., clustering), cycling, and reciprocity; an important attribute of the model because such dependence features characterize valued networks \cite{Wyatt:2010}.

There are two specification steps in our approach to GERGMs: first, we specify a tractable joint distribution that captures the dependencies of interest on a restricted network, $X$, and then we transform $X$ onto the support of $Y$; thus producing a probability model for $Y$.
To illustrate these steps, begin with consideration of the restricted valued network $X \in [0,1]^m$, where $m$ is the number of edges.

In our first specification step, $\bh$ is formulated to represent joint features of $Y$  in the distribution of $X$:
\begin{equation}
 f_X(X,\bt) = \frac{\exp \left[\bt' \bh(X)\right]}{\int_{[0,1]^m} \exp\left[\bt' \bh(Z)\right]dZ},  
 \end{equation} 
where $\bt \in \mathbb{R}^p$ is the parameter vector, $\bh:[0,1]^m \rightarrow \mathbb{R}^p$, $\bh$ is finite on $[0,1]^m$ and $h_i(\cdot)$ are the sums of subgraph products such that for every  $i,~\frac{\partial^2 \bh(X)}{\partial^2 X_{ij}} = 0$. This is a flexible specification because many dependence relationships can be captured by summing products over subgraphs of the network, particularly when the edges are in the unit interval\cite{Wyatt:2010}.  For instance, networks generated by a highly reciprocal process are likely to exhibit high values of $\sum_{i < j}X_{ij}X_{ji}$, and those in which connections gravitate toward high-degree vertices exhibit high values of $\sum_{i}\sum_{j,k \neq i}X_{ji}X_{ki}$ (i.e., ``two-stars'' \cite{Park:2004b}). An important property of $f_X$ is that when $\bt = \bm{0}$, $X$ is a network of independent uniform random variables.

In our second specification step, we apply parameterized, one-to-one, monotone increasing transformations ($G^{-1}(\cdot)$) to the $m$ edges of the restricted network, thus transforming the restricted network $X$ onto the support of the network of interest $Y$. $Y_{ij} = G^{-1}_{ij}(X_{ij}, \bl_{ij})$, where $\bm{\lambda}_{ij}$ parameterizes the transformation to capture marginal features of $Y_{ij}$. 
Because $dG^{-1}(X_{ij},\bm{\lambda}_i)/d X_{ij} > 0$, the properties of multivariate transformations\cite{Casella:2001} imply that the distribution of $Y$ is 
$  f_Y(Y, \bt, \bm{\Lambda}) = f_X(\bm{G}(Y,\bL),\bt)|J|, $
 where the Jacobian matrix, $J$, is the matrix of first partial derivatives.  Since $J$ is a diagonal matrix, we may write the GERGM as 
\begin{equation}  
f_Y(Y ,\bt, \bm{\Lambda}) = \frac{\exp \left[\bt' \bh(\bm{G}(Y,\bm{\Lambda}))\right]}{\int_{[0,1]^m} \exp\left[\bt' \bh(Z)\right]dZ}\prod_{ij}g(Y_{ij},\bm{\lambda}_{ij}). 
\label{ll} 
\end{equation}

A useful way to specify $g$ is as a probability density function (i.e., $G$ is a CDF, and $G^{-1}$ an inverse CDF) parameterized to match the support of $Y$ and capture features of $Y$ such as location, scale, and dependence on covariates. 
This approach to specifying $g$ has the elegant feature that the distribution contains many common models for independent and identically distributed variables as special cases when $\bt = \bm{0}$. 
For instance, if $g$ is a Gaussian PDF with constant variance and the mean dependent on a vector of covariates, the model reduces to that assumed in least squares regression. 
The GERGM also allows hypothesis tests for block restrictions (i.e., likelihood ratio or Wald tests) to test the assumption that the edges of $Y$ are independent conditional upon $\bm{\Lambda}$.

\begin{figure}[htp]
\begin{center}
\begin{tabular}{cc}
 {\large $f_X$} & {\large $g(Y)$}\\
 \includegraphics[scale=.55]{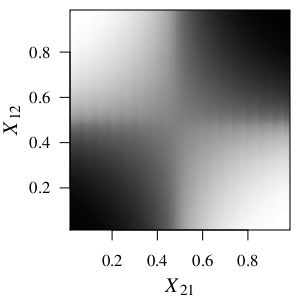} & \includegraphics[scale=.55]{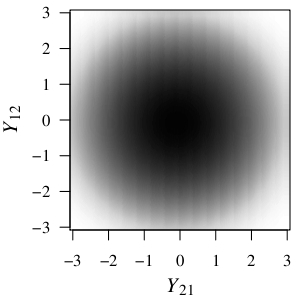} \\
\end{tabular}
\end{center}
\vspace{-.6cm}
{\large $f_Y$} \\
\includegraphics[scale=.6]{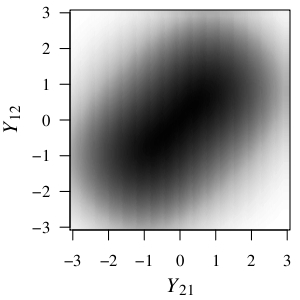}
\caption{Bivariate distributions for edges in a two-vertex di-graph. The darker the shading, the higher the relative likelihood of a point. In this example, $g$ is the standard normal PDF, and $f_X$ is defined by $\bh = \{X_{12}+X_{21},X_{12}X_{21}\}$, and $\bt = \{-3.5,7\}$, representing negative density and positive reciprocity effects.}
\label{fgf}
\end{figure}

There are two ways to interpret dependence modeling of $Y$ via $X$. First, following \cite{Wyatt:2010}, who derive an ERGM-like model for a network with discrete edges on the unit interval, $X$ can be interpreted as a standardized relational intensity network. Second, and more directly, when $g$ is a PDF, $X$ is the random variable drawn from the joint distribution of the quantiles of $Y$. Therefore, the vectors $\bh$ and $\bt$ characterize the dependencies among the quantiles of $Y$.  The latter interpretation closely resembles the process of constructing joint distributions with copula functions \cite{Gleeson:2008,Sato:2010}. A simple example of deriving a joint distribution through the combination of $\bh$ and $g$ is illustrated in figure \ref{fgf}, which presents the distributions of $X$ and $Y$ for a directed network with two vertices exhibiting a high degree of reciprocity. 

Estimation of the parameters in the model is a non-trivial task. The greatest challenge in estimating $\bt$ and $\bL$ in equation \ref{ll} is that the integral in the denominator is typically intractable. Because of the polynomial structure of ${\text {\bf h}}$, and the fact that the variables of integration are bounded, we know that the integral is both positive and finite, meaning $f_Y$ is a proper joint distribution. However, inference requires the approximation of the denominator. 

In order to approximate the denominator in equation \ref{ll}, we sample from $f_X$ using a Gibbs Sampler. To do so, we require the conditional distribution of $X_{ij}|X_{-ij}$. To simplify the notation, let $\int_{[0,1]^m} \exp\left[\bt' \bh(Z)\right]dZ = C(\bt)$. The conditional distribution ($f^c_{X}$) is given by 
\begin{eqnarray} 
 f_X^c(X_{ij} |\bt) & =& \frac{\exp\left[ X_{ij} \bt' \frac{\partial \bh(X)}{\partial X_{ij}}\right]}{\bt' \left(\frac{\partial \bh(X)}{\partial X_{ij}}\right)^{-1}\left[ \exp(\bt' \frac{\partial \bh(X)}{\partial X_{ij}})-1\right]} .
  \label{cd} 
 \end{eqnarray}
We may then draw from the conditional distribution in equation \ref{cd} using the inverse CDF method. If $u$ is a uniform (0,1) random variable, then 
\begin{equation}
X_{ij}|X_{-ij} \sim \frac{ \ln \left[1+u\left(\exp \left[ \bt' \frac{\partial \bh(X)}{\partial X_{ij}} \right]-1\right)\right]}{ \bt' \frac{\partial \bh(X)}{\partial X_{ij}}}.
\end{equation}
When  $\bt' \frac{\partial \bh(X)}{\partial X_{ij}} = 0$ the conditional density given in equation \ref{cd} is undefined. However, in this case, each point in the unit interval is equally likely and the conditional distribution of $X_{ij}$ is uniform(0,1).

In order to estimate $\bt$ and $\bL$, we maximize $\ln\left[f_Y\right]$: 
\begin{equation} 
\bt' \bh(\bm{G}(Y,\bm{\Lambda})) + \sum_{ij}\ln \left[g(Y_{ij}|\bl_{ij})\right]-\ln\left[C(\bt)\right].
\label{logl}
\end{equation}
Our algorithm iteratively proceeds by maximum likelihood (ML) estimation of $\bL|\bt$ and Markov chain Monte Carlo maximum likelihood estimation (MCMC-MLE) of  $\bt|\bL$ until convergence. We derive an approximation to the asymptotic variance-covariance matrix by the inverse of the negative Hessian matrix at the last iteration. 

The estimation of $\bL|\bt$ is straightforward. Because $C(\bt)$ does not depend on $\bL$, ML estimation of $\bL|\bt$ reduces to
\begin{equation}
\argmax_{\bL} \left( \bt' \bh(\bm{G}(Y,\bm{\Lambda})) + \sum_{ij}\ln \left[g(Y_{ij}|\bl_{ij})\right] \right),
\end{equation}
a function easy to maximize using a hill-climbing algorithm.

The estimation of $\bt|\bL$ is more involved. Let $\widehat{X} = \bm{G}(Y,\hat{\bL})$ be the estimate of the intensity/quantile network given the current estimate of the transformation parameters. The second term in equation \ref{logl} does not depend on $\bt$, so to estimate $\bt|\bL$ we find 
\begin{equation}
\argmax_{\bt} \left(\bt' \bh(\widehat{X}) -\ln\left[C(\bt)\right] \right),
\end{equation}
which requires an approximation of $C(\bt)$. We approximate $C(\bt)$ using MCMC-MLE; an iterative method itself. Let $\bt^{[i-1]}$ be the previous estimate of $\bt$, and $\tilde{\bm{X}}$ be a sample of $n$ networks drawn from $f_X(X,\bt^{[i-1]})$.  Then, an approximation to $C(\bt)$ is given by 
\begin{equation}
\widehat{C(\bt)}  = C(\bt^{[i-1]})\sum_{j=1}^n \frac{\exp \left[\bt' \bh(\tilde{X}_j)\right]}{\exp \left[\bt'^{[i-1]} \bh(\tilde{X}_j)\right]}.
\end{equation}
This requires a starting value for $\bt$. In simulation experiments, we have found the pseudolikelihood estimate ($\argmax_{\bt}\left(\sum_{ij} \ln\left[ f^c_{X}(X_{ij} | \bt )\right] \right)$) to be effective in providing starting values for $\bt$ (i.e., $\bt^{[0]}$).

\begin{figure}[htp]
\begin{center}
(a) Regression Estimates \\ 
\includegraphics[scale=1]{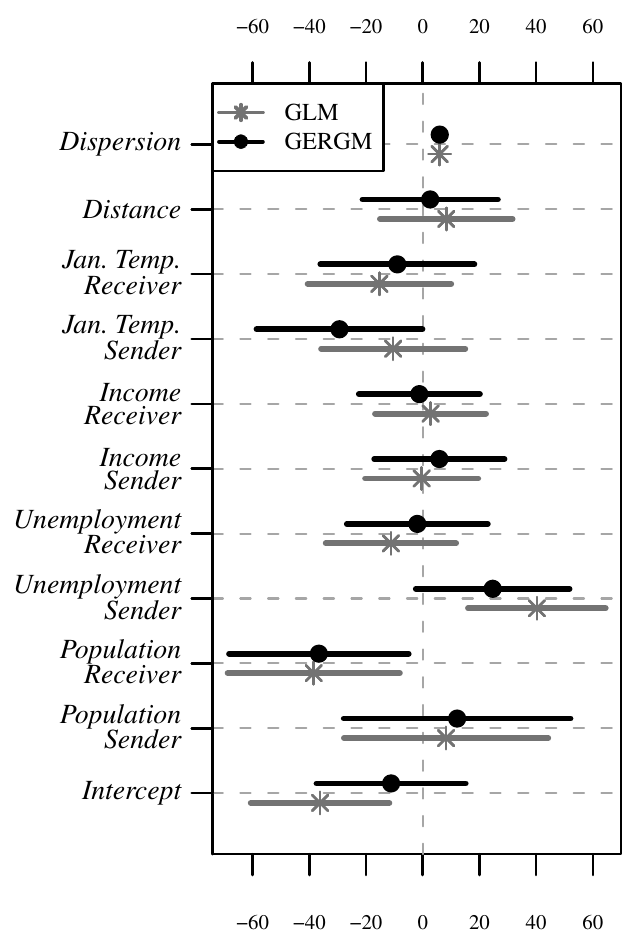} \\
(b) Dependence Estimates \\
 \includegraphics[ scale=1]{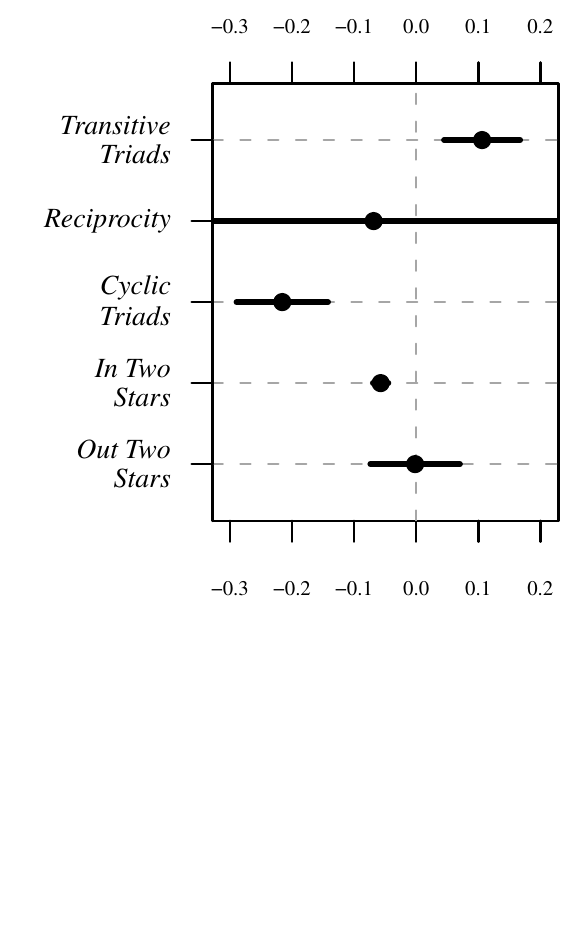}
\end{center}
\label{results}
\vspace{-3.5cm}
\caption{Estimates of the parameters: bars span 95\% confidence intervals. 5,000 draws for three iterations used in the MCMC-MLE}
\label{results}
\end{figure}

\begin{figure}[htp]
\begin{center}
\begin{tabular}{cc}
(a) Cycles & (b) Dyadic Reciprocation\\
 \includegraphics[scale=.6]{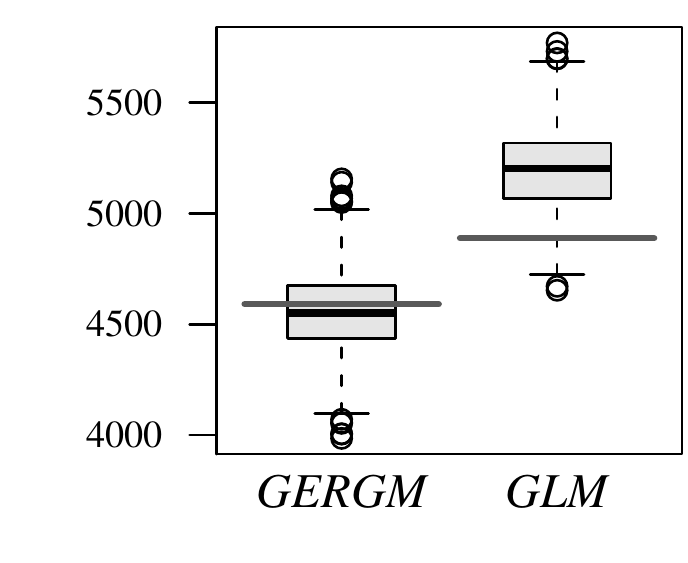} & \includegraphics[scale=.6]{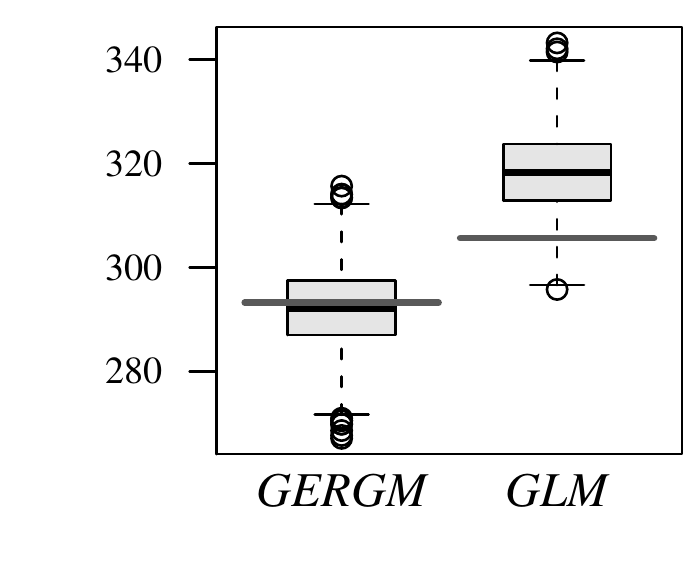} \\
\end{tabular}
\end{center}
\vspace{-.5cm}
\caption{Reciprocal Feature Prediction: The boxplots represent the respective dependence statistic computed on 1,000 instances of the latent intensity network drawn from each model. Let $\widehat{X}$ be the respective estimate of the intensity network obtained as the CDF evaluated at the transormation parameters ($\bL$) for the GERGM and Cauchy GLM. Then cycles (a) is $\sum_{i <j <k} \widehat{X}_{ij}\widehat{X}_{jk}\widehat{X}_{ki} + \widehat{X}_{ik} \widehat{X}_{kj}\widehat{X}_{ji}$, and dyad reciprocation (b) is $\sum_{i < j} \widehat{X}_{ij}\widehat{X}_{ji}$. Horizontal grey bars are placed at the statistic computed on the estimated intensity network.}
\label{boxes}
\end{figure}

We illustrate important features of the GERGM and demonstrate its efficacy by applying it to a real-world network: domestic migration in the United States\cite{Ke:2006,Ke:2006b}. We model changes in the directional migration flows between the 50 United States (as well as Washington D.C. and Puerto Rico) between 2006 and 2007. $Y_{ij}$ is the difference between the number of people who migrated from state $i$ to state $j$ in 2007 and the number who migrated from $i$ to $j$ in 2006.
These data allow us to consider the GERGM in the context of a valued network requiring transformation away from an intensity network onto a continuous unbounded support with exogenous covariates and endogenous parameters, thus making full use of the GERGM's flexibility.
We use the Cauchy distribution as our $g$ function because its thick tails capture the high empirical kurtosis (637) of the network \cite{Mizera:2002}. Thus, in the case where the edges of the network are independent conditional on the covariates, this specification reduces to a generalized linear model (GLM) \cite{Nelder:1972} with a Cauchy link function.
Because previous work on interstate migration\cite{Chun:2008} suggests that population, unemployment, per-capita income, and mean January temperature of both the sending and receiving states are significant determinants of migration, we include the change in each of these variables from 2005 to 2006 as covariates in our GERGM. 
We complete our specification by including endogenous dependence terms for clustering, dyadic reciprocity, generalized reciprocity (i.e., cycling -- the degree to which change in flows to and from a state are correlated\cite{Jian:2008}), state level attraction, and state level repellence.

Figure~\ref{results} shows the estimates from our GERGM as well as estimates from a Cauchy GLM. 
A Wald test suggests the restriction of the dependence terms to zero in the regression model is inappropriate and that the GERGM provides a better fit to the data (Wald statistic $=$ 119.19 on 5 degrees of freedom, statistically significant at the 0.001 level).
The statistically significant effects for the network parameters indicate that (a) there are clustering effects in the network, (b) migration to states repels further migration, and (c) increases in migration flows from a state are not offset by increases in flows to that state.
We also find a decrease in the number of people leaving warm states, a decrease in migration to states that experienced a substantial increase in population in the previous year, and evidence of an increase in migration away from states experiencing increases in unemployment.

The superior performance of the GERGM relative to the Cauchy regression is further depicted in figure \ref{boxes}, which gives the predicted and observed network-level reciprocity and cycling measures from the GERGM and Cauchy GLM. This figure shows that the regression does not adequately fit the lack of reciprocity in the migration network. Theoretically, it is expected that a network of change in migration would exhibit anti-reciprocity and anti-cycling. If a locale is experienceing a spike in migration to other places, that is likely indicative of some undesireable feature of said locale. This anti-reciprocal feature of the migration network cannot be integrated into the conventional regression modeling framework.

Our GERGM model greatly expands the scope of networks which can be modeled within the ERGM framework. 
We used this technology to analyze a real-world network and produce insights that could not be produced without the GERGM. 
Our general model represents a major advance in the statistical analysis of networks, and we expect it to become a common tool in disciplines spanning the sciences.   

\vspace{0.4cm}

\noindent \emph{The authors thank James Fowler and Peter Mucha for useful comments. This work was supported in part by a grant from the University of Massachusetts Amherst College of Social and Behavioral Sciences.}

\bibliographystyle{apsrev4-1}
\bibliography{gergm.bib}

\end{document}